\documentclass[11pt,a4paper]{article}
\usepackage[T1]{fontenc}
\usepackage[utf8]{inputenc}
\usepackage{lmodern}
\usepackage{amsmath,amssymb,graphicx,booktabs}
\usepackage[margin=2.6cm]{geometry}
\usepackage[hidelinks]{hyperref}
\usepackage{microtype}
\title{Numerical verification limits of asymmetric intrastromal hydrogel reinforcement for decentered keratoconus}
\author{J. Sumaya-Mart\'{i}nez\thanks{Faculty of Sciences, Universidad Aut\'{o}noma del Estado de M\'{e}xico, Toluca, Mexico} \and Alan Altamirano-Torres\thanks{Centro de Microcirug\'{i}a Ocular (CMO), Mexico}}
\date{1 November 2026}
\begin{document}
\maketitle
\begin{abstract}
Spatially tailored intrastromal hydrogel scaffolds have been proposed for decentered keratoconus, but their mechanical and optical benefits require stringent numerical validation. We reanalyze conforming two-material finite-element benchmarks and earlier optomechanical screening calculations, distinguishing solver verification from constitutive adequacy and treatment efficacy. In a curved corneal-patch benchmark with a fixed-volume rectangular inclusion, the apparent peak-displacement benefit of an offset over a centered inclusion changes from $-5.66\%$ to $+1.73\%$ across four tetrahedral meshes, while absolute displacements remain strongly mesh dependent. On the finest mesh, the quadratic term in the Green--Lagrange strain tensor contributes about 18\% of the root-mean-square magnitude of the linearized strain; the underlying finite-element solutions themselves remain linear elastic. Earlier optical screening was sensitive to corneal orientation, mesh refinement and wavefront-reference conventions and is not retained as validated evidence. These observations do not establish superiority of asymmetric reinforcement. Instead, they motivate prespecified criteria for geometry-invariant finite-strain simulations, two-material interface verification, optical wavefront convergence, and independent experimental validation. No clinical recommendation is implied.
\end{abstract}
\noindent\textbf{Keywords:} keratoconus; corneal biomechanics; hydrogel; finite-element verification; optical wavefront; grid convergence.
\medskip
\section{Introduction}
Intrastromal reinforcement for decentered keratoconus poses two distinct design questions: whether a scaffold redistributes local mechanical strain, and whether the resulting change in both corneal surfaces improves the transmitted optical wavefront. A favorable answer to one does not establish a favorable answer to the other. Finite-element studies of intrastromal ring segments already show that implant geometry, corneal material response, and individual morphology matter \cite{lago2015,kling2013,ring2021,ring2023}. Liu and colleagues recently combined a concave, multi-crosslinked GelMA-based scaffold with finite-element analysis and rabbit-model observations \cite{liu2026}. Consequently, ``hydrogel reinforcement'' or a nonuniform implant shape alone cannot establish novelty. The contribution to be tested is a \emph{quantitatively robust, volume-matched optomechanical advantage} for a prescribed decentered ectasia.

A risk in computational design is optimizing an apparent improvement that is smaller than discretization error or depends on a nonphysical optical reference. Here we revisit the existing simulations and report verification results honestly, even where they weaken the design hypothesis. We distinguish the independently executed conforming two-material benchmark from earlier nonlinear curved-cap screening, because those models use different materials, domains, boundary conditions and optical conventions. Their numerical outputs must not be pooled as if they formed a single mesh sequence.

\subsection{Scope and scientific question}
The study is deliberately narrower than a device-performance claim. Its testable question is whether an apparently favorable positional change in an intrastromal inclusion can be distinguished from changes in the finite-element discretization, chosen observable, and constitutive approximation. This distinction matters for a decentered ectasia, where a change in local stress need not translate into a beneficial change in the transmitted wavefront. A displaced implant may reduce deformation in one portion of the cornea while increasing it in another; a single global displacement maximum cannot reveal that redistribution.

Three levels of evidence are kept separate. \emph{Code verification} concerns the faithful numerical solution of a defined mathematical problem. \emph{Model validation} concerns agreement between the chosen mechanics, optical conventions, and independent physical observations. \emph{Treatment efficacy} would require biological and clinical evidence that is not supplied by any of the calculations reported here. The absence of the latter is not remedied by a small residual norm or by a graph that appears visually smooth. This distinction also constrains the title and conclusions of the present preprint.

\subsection{Why optical and mechanical objectives cannot be substituted}
A structural response is a vector field $\mathbf u(\mathbf X)$ over a three-dimensional domain, while a transmitted wavefront is an optical-path function over an entrance or exit pupil. An RMS displacement integrates different information from the coefficient of a third-order pupil polynomial. Furthermore, coma is affected by the direction and position of rays leaving both surfaces, so an anterior-surface height correction alone is not a wavefront prediction. For these reasons, a numerical design that optimizes a mechanical endpoint must be examined optically independently, with material and reference-state uncertainties propagated through the computation.

\section{Model and verification methods}
\subsection{Conforming two-material benchmark}
The latest independent benchmark is a curved square corneal patch, not a whole eye and not a patient-specific tomography. A rectangular intrastromal inclusion with dimensions $1.4\times1.4\times0.10\,\mathrm{mm}$ occupies a fixed theoretical volume of $0.196\,\mathrm{mm}^3$. The centered and shifted versions use the same shape, material constants, and discretized interface topology. Nodal continuity on the shared interface represents perfect bonding, not measured interfacial adhesion. A linear isotropic elastic constitutive law, a prescribed pressure load on the reference surface, and clamped patch edges were used. Consequently this model is a \emph{numerical diagnostic}, not a physiological implantation simulation.

The test compares four refinements with 162, 1,296, 4,374 and 10,368 tetrahedra. For each solution we report maximum displacement and an independently defined mean central axial displacement, rather than selecting the more favorable observable after seeing the results. The geometry of the material inclusion is fixed, although the faceted approximation of the curved external surface changes with refinement.
\subsection{Postprocessing for finite-strain adequacy}
For each piecewise-linear tetrahedral displacement field we compute $\mathbf G=\nabla\mathbf u$, $\mathbf F=\mathbf I+\mathbf G$, the infinitesimal strain $\boldsymbol\varepsilon=(\mathbf G+\mathbf G^T)/2$, and the Green--Lagrange strain
\begin{equation}
\mathbf E=\boldsymbol\varepsilon+\frac12\mathbf G^T\mathbf G.
\end{equation}
The volume-weighted RMS magnitude of the quadratic contribution divided by the RMS magnitude of $\boldsymbol\varepsilon$ is a \emph{kinematic nonlinearity diagnostic}; it is neither a relative displacement error nor proof that the linearly computed field satisfies nonlinear equilibrium. Positive $\det\mathbf F$ tests local orientation of this computed field, not the validity of its underlying material law.
\subsection{Verification, validation, and optical criteria}
Previous calculations separately verified a manufactured-solution linear two-material FEM, a curved follower-pressure operator and its virtual-work derivative, and elementary two-interface Snell refraction. These component checks do not automatically verify the fully coupled corneal problem. In addition, earlier optomechanical screening changed apparent coma improvement with mesh density, pupil size, Zernike truncation and optical-surface orientation. Optical results based on those inconsistent geometries are excluded from the primary results reported here.

A future coupled assessment must use the same physical anterior/posterior surfaces and implant subdomain across refinements, reconstruct the physiological loaded reference state, and couple a validated pressure tangent to a hyperelastic anisotropic stroma and independently characterized hydrated hydrogel. The reference sphere and exit pupil must be determined by the optical model, and the Zernike basis and pupil diameter fixed before comparison. The two prespecified primary endpoints are (i) volume- or area-weighted 95th-percentile principal Green--Lagrange strain in a fixed ectatic region of interest and (ii) two-surface ray-traced third-order coma RMS. Peak displacement, interface traction, total higher-order RMS and optical image metrics are secondary endpoints.

\subsection{Continuum equations and interface conditions}
Let the reference patch occupy $\Omega_0=\Omega_s\cup\Omega_h$, where $\Omega_s$ and $\Omega_h$ denote stromal and hydrogel domains. On the internal interface $\Gamma_{sh}$, the benchmark enforces perfect kinematic bonding and balance of traction,
\begin{equation}
[\![\mathbf u]\!]=\mathbf 0, \qquad [\![\mathbf P\mathbf N]\!]=\mathbf 0\quad\text{on }\Gamma_{sh},
\label{eq:interface}
\end{equation}
where $\mathbf P$ is the first Piola--Kirchhoff stress and $\mathbf N$ is a reference-surface normal. These statements specify the intended continuum model; the linear benchmark uses the corresponding small-strain stress $\boldsymbol\sigma=\mathbb C:\boldsymbol\varepsilon$. Since the same finite-element nodes are shared across the interface, displacement continuity is imposed strongly. This is not a numerical representation of hydrogel delamination, friction, imperfect adhesion, or degradation.

For a geometrically nonlinear future model, the governing equilibrium in the reference domain is
\begin{equation}
\operatorname{Div}\mathbf P+\rho_0\mathbf b=\mathbf 0,\qquad
\mathbf F=\mathbf I+\nabla_{\!X}\mathbf u,\qquad J=\det\mathbf F>0.
\label{eq:equilibrium}
\end{equation}
The weak form for an admissible virtual displacement $\mathbf v$ satisfying the essential constraints is
\begin{equation}
\int_{\Omega_0}\mathbf P:\nabla_{\!X}\mathbf v\,dV
-\int_{\Gamma_{p,0}}p\,J\mathbf F^{-T}\mathbf N\cdot\mathbf v\,dA
-\int_{\Omega_0}\rho_0\mathbf b\cdot\mathbf v\,dV=0.
\label{eq:weak}
\end{equation}
The pressure term follows from Nanson's relation and is a \emph{follower} load. Equation~\eqref{eq:weak} is a specification for the decisive nonlinear calculation, not a claim that the conforming rectangular-inclusion benchmark already used this weak form. In that benchmark, the applied pressure was defined on the reference surface and the material response was linear elastic. Mixing the two formulations in a purported single convergence sequence would be invalid.

\subsection{Finite-strain constitutive requirements}
For illustration, a compressible neo-Hookean strain-energy density could be written
\begin{equation}
W(\mathbf F)=\frac{\mu}{2}\bigl(J^{-2/3}I_1-3\bigr)
+\frac{K}{2}(J-1)^2,\qquad I_1=\operatorname{tr}(\mathbf F^T\mathbf F),
\label{eq:nh}
\end{equation}
with $\mathbf P=\partial W/\partial\mathbf F$. This law supplies a basic large-deformation test, but the cornea is anisotropic, lamellar and depth-dependent; a patient-specific prediction would require an independently calibrated material description, such as a validated fiber-reinforced formulation. Similarly, hydrogel stiffness is expected to depend on composition, hydration, temperature and fabrication. Assigning an illustrative modulus without identifying it experimentally is useful for sensitivity studies only. No parameter-fitting outcome is claimed here.

\subsection{Finite-element consistency and geometry control}
A continuous physical inclusion is held fixed in the benchmark: $V_h=(1.4)(1.4)(0.10)=0.196\,\mathrm{mm}^3$, independent of the number of tetrahedra. This is distinct from selecting a new set of elements after each refinement based on element-centroid location, which can change the physical design being compared. Interface conformity is a necessary but not sufficient condition for convergence: external curved boundaries, clamp locations, pressure faces, element aspect ratios and quadrature also require consistent refinement.

The benchmark uses nodally continuous, piecewise-linear tetrahedral displacement fields. Its discrete equilibrium is a linear system $\mathbf K_h\mathbf u_h=\mathbf f_h$ with shared interface degrees of freedom. Residuals document whether the assembled algebraic equations were solved. They do not measure the error $\mathbf u_h-\mathbf u_{\rm exact}$, nor guarantee that a particular derived quantity has converged. A fixed physical point, rather than an arbitrary nearest mesh node, should be used for pointwise comparisons wherever possible.

\subsection{Prespecified mechanical functionals}
For a fixed region of interest $\Omega_e$ containing the ectatic zone, a useful future finite-strain observable is the volume-weighted 95th percentile of the largest principal value of $\mathbf E$. Other potentially relevant functionals include the spatial distribution of interfacial normal and shear traction, maximum principal stress in the adjacent stroma, surface curvature, and the deformation of the optical zone. These quantities should be reported with signs, locations, and uncertainty, not compressed into a single favorable percentage.

The present benchmark reports two simpler, \emph{non-equivalent} functionals: a maximum displacement $D_\infty$ over the full model and a mean axial displacement $D_c$ in a predefined central region. For any such scalar endpoint $D$, define the centered-to-shifted contrast
\begin{equation}
A_D(h)=100\,\frac{D_{\rm centered}(h)-D_{\rm shifted}(h)}{D_{\rm centered}(h)}.
\label{eq:advantage}
\end{equation}
Positive $A_D$ favors the shifted inclusion for that endpoint alone. It is not a percentage reduction in disease severity, mechanical failure probability, or wavefront aberration.

\section{Results}
\subsection{Grid dependence and endpoint disagreement}
Table~\ref{tab:mesh} reports the reproduced two-material linear benchmark. With increasing refinement the sign and magnitude of the apparent benefit change. At 1,296 tetrahedra the shifted inclusion improves the maximum-displacement endpoint but worsens the mean central axial displacement. The two finest grids change the centered maximum displacement from $239.80$ to $287.00\,\mu\mathrm{m}$ (19.69\%) and the shifted maximum from $234.38$ to $282.04\,\mu\mathrm{m}$ (20.33\%). Such drift rules out an independent-of-mesh effect size.
\begin{table}[htbp]\centering
\caption{Two-material conforming benchmark, NOT the nonlinear corneal scaffold model. Positive advantage denotes a lower displacement for the shifted inclusion. The volume-matched rectangular inclusion is identical across models.}
\label{tab:mesh}
\begin{tabular}{rrrrr}\toprule
Tetrahedra & Centered ($\mu$m) & Shifted ($\mu$m) & Peak advantage & Central advantage\\\midrule
162 & 84.54 & 89.33 & $-5.66\%$ & $-4.42\%$\\
1296 & 176.17 & 171.83 & $+2.46\%$ & $-3.28\%$\\
4374 & 239.80 & 234.38 & $+2.26\%$ & $+1.22\%$\\
10368 & 287.00 & 282.04 & $+1.73\%$ & $+2.43\%$\\\bottomrule
\end{tabular}\end{table}
\begin{figure}[htbp]\centering\includegraphics[width=.86\linewidth]{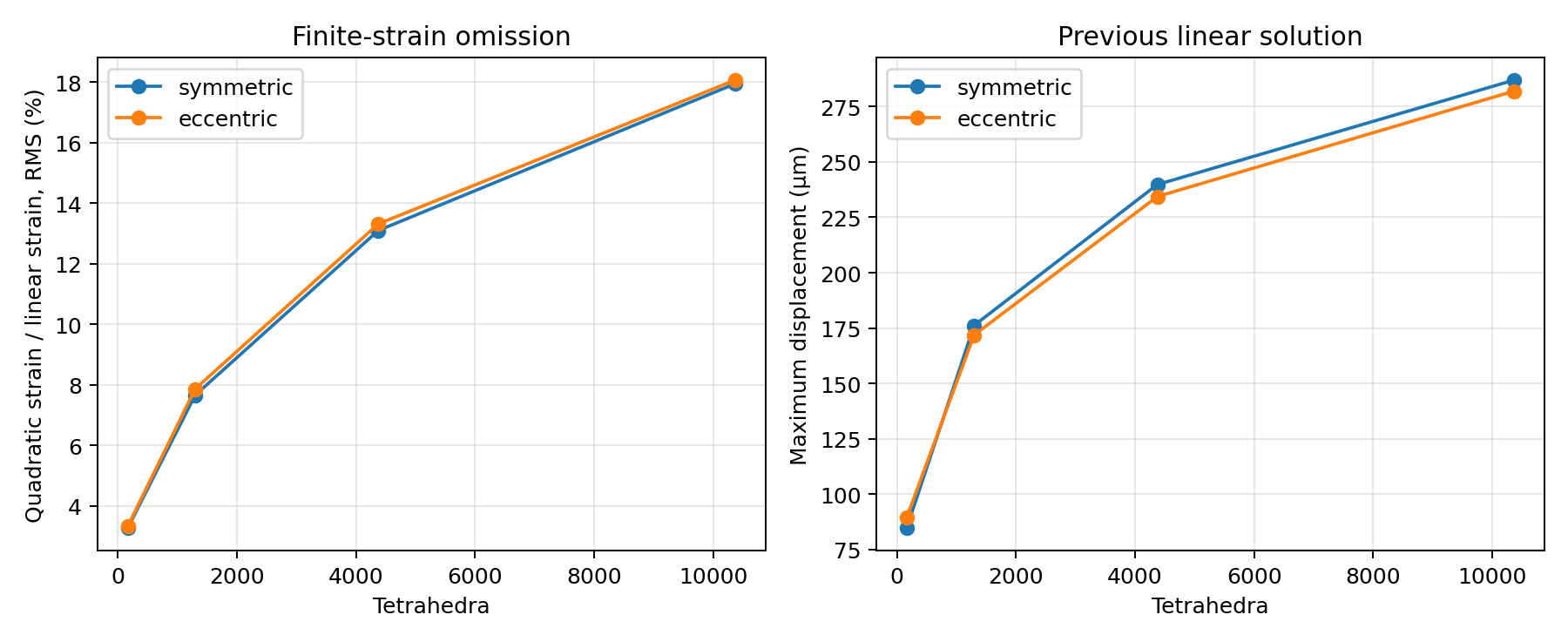}
\caption{Independently recomputed finite-kinematics diagnostic on the linearly solved, conforming two-material benchmark. This graphic does not represent a nonlinear finite-element equilibrium solution.}\label{fig:kinematics}\end{figure}
\subsection{Finite-kinematics diagnostic}
For 10,368 tetrahedra, the centered and shifted benchmark fields have peak displacements of 287.00 and $282.04\,\mu\mathrm{m}$, respectively. The quadratic kinematic term is 17.95\% and 18.07\% of the volume-weighted RMS linear-strain magnitude. The maximum $\|\nabla\mathbf u\|_F$ values are 0.190 and 0.223; the minimum $\det\mathbf F$ values are 0.961 and 0.949. These results show that a linear small-strain model is not an adequate basis for inferring a physiological 1--2\% advantage, but do not quantify the displacement error from nonlinearity.
\subsection{Status of previous optomechanical calculations}
An earlier neo-Hookean corneal-cap model was solved with effective intrastromal material masks and a follower pressure. Across development iterations, changes in the implant representation, the corneal optical orientation, and the numerical pupil sampling materially altered the predicted relative coma benefit; in a convex-geometry recalculation it changed sign on the finest available mesh. Subsequent pressure-direction checks were also found to require careful matching of material-face labels to the optical convention. These are \emph{development diagnostics}, not independent evidence of efficacy. The previously advertised reductions in coma or peak displacement therefore are not carried forward as validated study outcomes.

\subsection{Quantitative resolution sensitivity}
The 10,368-element values differ from their 4,374-element counterparts by $47.20\,\mu\mathrm m$ for the centered inclusion and $47.66\,\mu\mathrm m$ for the shifted inclusion. Relative to the values on the coarser of these two grids, these changes are approximately $19.69\%$ and $20.33\%$, respectively. The $+1.73\%$ finest-grid advantage is therefore not separated from the grid sensitivity of the absolute displacement fields. This comparison is a warning sign, not a formal upper error bound: two-grid differences can underestimate or overestimate the exact-solution error and need not represent the asymptotic convergence regime.

The number of elements increases but the observable does not approach an apparent plateau. It would be scientifically misleading to perform a Richardson extrapolation, claim a grid-convergence index, or fit an arbitrary empirical order to these values without additional resolution levels that establish an asymptotic regime. Nor can a model change between meshes be treated as a discretization-only perturbation. In a decisive study, both the integral norms of the full displacement field and the predefined clinical/mechanical functionals should converge, with a tolerance smaller than the claimed between-design difference.

\subsection{The effect of endpoint selection}
On the 1,296-element grid, $A_{D_\infty}=+2.46\%$ while the central axial-displacement contrast is $-3.28\%$. These signs are not contradictory: a maximum over a large domain and a mean over its center weight spatial variation differently. At the coarsest grid both contrasts are negative; at the two finest grids both are positive, but their magnitude still varies. A credible design ranking must thus specify the optimization objective before computing the solutions, report spatial strain and stress maps rather than only maxima, and test whether improvements persist when clinically meaningful constraints are imposed.

\subsection{What the kinematic diagnostic does and does not establish}
The Green--Lagrange expansion in the preceding section is an exact algebraic identity for a given deformation gradient. The quadratic term can therefore be evaluated even if the displacement field came from a linearly solved problem. Its roughly 18\% relative magnitude on the finest mesh indicates that ignoring finite rotations and stretches can modify the interpreted strain measure substantially. However, it does not establish that a nonlinear finite-element solution would differ by 18\%, nor does it identify whether the centered or shifted design would be preferred after nonlinear re-equilibration. A fresh nonlinear solve, with an independently tested consistent tangent and load implementation, is required.

\subsection{Evidence hierarchy across the computational development}
Table~\ref{tab:evidence} separates the numerical component checks from the main design comparison. This hierarchy prevents a successful elementary test from being used to certify an unrelated part of the coupled model.
\begin{table}[htbp]\centering\small
\caption{Scope of the available computational evidence. ``Previously executed'' denotes a component or development test, not an independently validated patient-specific calculation.}
\label{tab:evidence}
\begin{tabular}{p{0.31\linewidth}p{0.26\linewidth}p{0.32\linewidth}}\toprule
Test & Established property & What it does not establish\\\midrule
Manufactured linear two-material elasticity benchmark & Implementation on a prescribed interface test problem & Nonlinear corneal deformation or measured hydrogel behavior\\
Follower-pressure directional derivative and virtual work & Operator-level consistency under specified boundary conditions & Correct reference configuration for the cornea\\
Planar two-interface refraction & Snell and optical path in a known geometry & Human-eye exit-pupil wavefront or image quality\\
Conforming curved-patch comparison & Same nominal inclusion volume with shared interface nodes & Converged deformation or clinically optimal placement\\
Finite-kinematics postprocessing & Size of omitted quadratic strain terms for a linear field & Nonlinear force equilibrium or clinical safety\\\bottomrule
\end{tabular}
\end{table}

\section{Discussion}
\subsection{What the numerical evidence supports}
The study supports the narrower claim that a fixed-volume, conforming two-material benchmark can be assembled and independently audited, and that its apparent positional advantage fails robustness checks at the tested resolution. A small effect that changes with the endpoint or mesh does not warrant a therapeutic interpretation. The larger quadratic kinematic term strengthens the case for a fully nonlinear re-solution, rather than postprocessing a linear field as though that operation corrected its mechanics.
\subsection{Novelty relative to prior work}
Liu et al. have already reported a topologically tailored GelMA corneal scaffold with in vitro characterization and rabbit-model observations \cite{liu2026}; Lago et al. have developed patient-specific corneal ring simulations compared with postoperative outcomes \cite{lago2015}. This manuscript presently adds a critical verification analysis and an explicit falsification criterion, but \emph{does not yet deliver} an experimentally validated novel hydrogel treatment. A future original research submission to this journal should show robust mechanical and optical effects under volume control and report actual swelling, transparency, viscoelasticity and cytocompatibility measurements \cite{yue2015,li2025}.
\subsection{Requirements for a decisive study}
The next study must (1) fix a physical corneal and implant geometry across refinements with conforming material domains, (2) reconstruct the stress-free and physiologically pressurized reference configurations, (3) identify stromal and hydrated-hydrogel finite-strain parameters from independent measurements, (4) incorporate a documented interface constitutive choice and swelling sensitivity, (5) demonstrate field- and endpoint-level grid convergence with tolerances smaller than the asserted effect, and (6) compare optical wavefront and image metrics against held-out tomographic/aberrometric data. An improvement should be rejected when it vanishes under any of these predefined checks. These requirements are part of the scientific conclusion, not a claim that the calculations were already performed.

\subsection{Uncertainty budget and failure modes}
The uncertainty in a derived mechanical or optical contrast is not a single scalar inferred from a residual norm. It can arise from the geometry and its representation, boundary constraints, constitutive parameters, discretization, nonlinear solver tolerances, and the definition of the endpoint. The uncertainties can be correlated: for example, the same insufficient surface discretization can affect both an anterior-surface slope and a ray-traced coma coefficient. A proposed workflow should vary these factors systematically rather than interpret agreement under one arbitrary choice as robustness.

At minimum, a convergence report should include a fixed physical domain, a family of conforming meshes, element-quality summaries, a norm of the displacement difference interpolated into a common comparison space, and a goal-oriented error check for each primary endpoint. The latter is particularly important when a difference of a few micrometers is used to justify an effect of only one or two percent. The smallest two-grid change should not be confused with a confidence interval or a verified bound.

\subsection{An explicit optical validation pathway}
The proposed optical observable is a wavefront $W(\rho,\theta)$ referred to a physically determined exit pupil and reference sphere. For a pupil of radius $a$, with normalized coordinates $(\rho,\theta)$, define a Zernike decomposition
\begin{equation}
W(\rho,\theta)=\sum_{n,m}c_n^m Z_n^m(\rho,\theta),\qquad
C_{3}=\sqrt{(c_3^{-1})^2+(c_3^{+1})^2}.
\label{eq:coma}
\end{equation}
The normalization of $Z_n^m$ must be stated so that $C_3$ has a consistent physical interpretation. Reference-sphere location, pupil registration, wavelength and removal of piston, tilt and defocus must be prespecified. A wavefront obtained by merely accumulating optical path to an arbitrary plane is not automatically equivalent to a calibrated aberrometer measurement. Successful recovery of a known spherical or planar wavefront tests the tracer, but not corneal predictive accuracy.

A defensible coupled sequence is therefore: calculate a converged deformation of both corneal surfaces; reconstruct smooth, orientation-consistent interfaces; trace rays through anterior and posterior refractive boundaries; locate a consistent optical reference; fit the \emph{same} Zernike basis on the \emph{same} physical pupil; and assess optical error with both ray-density and FEM mesh refinements. The primary comparison should be based on the signed change in coma RMS and on a defined image-quality metric, such as a polychromatic point-spread function or modulation-transfer function, with the assumptions behind each metric documented.

\subsection{Sensitivity to hydration, adhesion, and prestress}
A fabricated hydrogel may swell, relax, soften, or change refractive index after implantation. Such changes can make a nominally fixed-volume implant design physically different from its as-fabricated shape. An interface constrained to share nodes represents a perfectly bonded idealization and will generally suppress modes associated with slip or debonding. In addition, the physiological cornea is already pressurized before surgery; using a tomographic loaded shape as an unstressed reference can impose an artificial mechanical baseline. None of these limitations is addressed by increasing mesh density alone. Each requires an independent material, interfacial, or inverse-mechanics specification.

\subsection{Predefined go/no-go criteria for a subsequent design study}
We propose the following requirements \emph{before} reporting efficacy: (i) identical continuous material geometries and pressure faces across mesh families; (ii) convergence of local field norms and all primary endpoints; (iii) verification of finite-strain equilibrium, interface traction balance, and follower-pressure work; (iv) independent identification or bounded uncertainty ranges for hydrated stromal and hydrogel properties; (v) optical convergence that includes pupil and reference choices; and (vi) replication in geometries not used to select implant location. If a design effect is smaller than its numerical or parameter uncertainty, it should be reported as unresolved rather than positive or negative.

This protocol can be used to falsify rather than merely refine the original design hypothesis. In particular, finding that a centered and shifted inclusion are indistinguishable within verified uncertainty would be a meaningful result: it would establish that additional geometric complexity has not justified itself for the tested case. Conversely, an advantage that survives all these checks could support a subsequent biomaterials and preclinical investigation, but would still not by itself establish clinical utility.

\section{Conclusions}
On the available evidence, placing an intrastromal hydrogel inclusion asymmetrically has \emph{not} been shown to improve the biomechanics or optical quality of a decentered keratoconic cornea. The fixed-volume conforming benchmark remains sensitive to mesh density, endpoint choice and finite-strain kinematics. These findings define the limits of the existing simulation and a reproducible route toward a falsifiable optomechanical design study. No clinical treatment recommendation follows from these results.

\appendix
\section{Derivation of the kinematic diagnostic}
Let $\mathbf x=\mathbf X+\mathbf u(\mathbf X)$. Then $d\mathbf x=\mathbf F\,d\mathbf X$. The squared line-element change is $d\mathbf x\cdot d\mathbf x-d\mathbf X\cdot d\mathbf X=2\,d\mathbf X\cdot\mathbf E\,d\mathbf X$, where $\mathbf E=\tfrac12(\mathbf F^T\mathbf F-\mathbf I)$. With $\mathbf F=\mathbf I+\mathbf G$, algebra gives
\begin{equation}
2\mathbf E=\mathbf G+\mathbf G^T+\mathbf G^T\mathbf G.
\end{equation}
For tetrahedra $T$ with volumes $V_T$, define the diagnostic
\begin{equation}
\eta_{\rm geom}=\frac{\left[\sum_T V_T\|\tfrac12\mathbf G_T^T\mathbf G_T\|_F^2/\sum_T V_T\right]^{1/2}}{\left[\sum_T V_T\|\tfrac12(\mathbf G_T+\mathbf G_T^T)\|_F^2/\sum_T V_T\right]^{1/2}}.
\end{equation}
The reported finest-grid values of approximately $0.1795$ and $0.1807$ use this type of volume-weighted comparison. This dimensionless ratio characterizes the field already computed; it is not a nonlinear constitutive correction applied to the field.

\section{Audit protocol for an independent replication}
An independent replication should recover the exact geometry from code or structured mesh files, list units and material parameters, report pressure surface identification and traction sign, and document the essential boundary conditions before postprocessing. Separate repositories or immutable tagged archives are preferable when the study has undergone multiple changes in sign conventions, physical geometry, or numerical model class. The source files for the present preprint contain the article and its figure only; full executable datasets are not yet deposited in a permanent research archive. Therefore, while aggregate values are documented, a third-party bitwise replication from this source package alone is not claimed.

The verification record should distinguish four checks: (a) algebraic convergence at fixed $h$; (b) manufactured or analytic solution recovery for selected operators; (c) spatial convergence of the complete nonlinear equilibrium problem at fixed physical geometry; and (d) validation against independent experimental observations. Passing one check does not imply that the next has been passed. This hierarchy is necessary to interpret small differences between design variants without overstating the fidelity of the computational model.

\section*{Statements and availability}
This manuscript reports a computational reanalysis with no new human or animal experiments. All geometrical, material and optical conclusions are limited to the numerical test problems described herein. No clinical efficacy or safety claims are made. Code and input/output data are not included in this arXiv manuscript source package; reproducibility of numerical claims requires a separately deposited, versioned computational archive. Author contribution, competing-interest and funding statements should be finalized and agreed upon by both authors before public submission.

\end{document}